\documentstyle[aps,preprint]{revtex}
\begin{document}
%\draft
\author{\"O.\ Legeza and J.\ S\'olyom}
\title{Response of finite spin-S Heisenberg chains to local perturbations}
\address{Research Institute for Solid State Physics and Optics,
     H-1525 Budapest, P.\ O.\ Box 49, Hungary}
\date{\today}
\maketitle

\begin{abstract}
We consider the properties of finite isotropic antiferromagnetic  
Heisenberg chains with $S=1/2, 1, 3/2$ spins when a weak magnetic field is
applied on a few sites, using White's density matrix renormalization group 
(DMRG) method. For the $S=1$ chain there exists only one length scale in 
the system which determines the behavior of the 
one- and two-point correlation functions both around the local perturbation 
and near the free boundary. For the critical, half-odd-integer spin cases 
the exponent of the spin-spin correlation function was found to be $\eta=1$, 
and the exponent of the decay of the site magnetization around the
perturbed site is $x_m =\eta /2 $. Close to a free boundary, however, 
the behavior is completely different for $S=1/2$ and $S>1/2$. 
\end{abstract}
\pacs{PACS numbers: 05.30.-d,02.70.+d,75.10.Jm}
\bigskip
\narrowtext
\section{Introduction}

Nowadays quantum spin chains are a very active field of research both 
theoretically and experimentally\cite{haldane}. In early studies mostly 
the spectrum of magnetic excitations was considered. By now it is well 
known that chains described by the Heisenberg Hamiltonian with 
half-odd-integer spin have a gapless spectrum. In their critical
behavior they belong to the same universality class, irrespective of the
spin and are expected to be equivalent\cite{wzw} to that of the 
Wess-Zumino-Witten (WZW) model with topological coupling constant $k=1$.  
Thus, the value of the critical exponent for the algebraically decaying 
spin-spin correlation function is $\eta=1$, independently of the value of 
$S$. On the contrary, the low energy spectrum of isotropic antiferromagnetic 
chains with integer spins has a gap to the lowest excited state occurring at
the boundary of the Brillouin zone, $q=\pi/a$. This leads to the
exponential decay of the spin-spin correlation function with a short 
correlation length.

Since the samples always contain finite chain segments, the effect of the
inhomogeneity of the field distribution and the decay of the induced 
surface magnetization are of eminent interest for experimentalists. From 
a theoretical point of view a related problem, namely that of the effect 
of a locally applied boundary field was studied \cite{affleck7}, but for 
the spin-1/2 case only. 

More generally the question arises, how an external local 
perturbation will affect the behavior of finite or infinite spin chains. 
Our aim in this paper is to study the spatial dependence of the site 
magnetization and the spin-spin correlations in an antiferromagnetic 
spin-S Heisenberg chain when a small magnetic field is applied locally. 
For this purpose we will determine, using the density matrix 
renormalization group (DMRG) method \cite{white}, the ground state 
properties of isotropic spin-S finite Heisenberg chains with $S=1/2, 
1, 3/2$ spins in a finite local field.

The layout of the paper is as follows. In Sec. II we give a
short description of the results known for the $S=1/2, 1, 3/2$ systems.
The model and a few main points of the numerical procedure 
applied in the course of the calculations are briefly mentioned in Sec III. 
The numerical results are presented in Sec. IV. Finally Sec V. contains a 
brief summary.
         
\section{Results known for the spin-S Heisenberg chains}

It is known from analytical \cite{vbs} and numerical \cite{white,szam,s1cikk} 
studies devoted to integer spin systems that due to the finite value of the 
energy gap the ground-state correlation functions fall off exponentially. 
The most accurate value for the correlation length of the two-point
correlation function $\langle S_i^z S_{i+l}^z \rangle$ was determined by 
White and Huse\cite{s1cikk}. This calculation gave $\xi_{\rm bulk}\simeq 
6.03(1)$ for the $S=1$ Heisenberg chain.
 
They have also studied the decay length of the local magnetization
near the surface. According to the valence bond model \cite{vbs} of
spin-1 chains in an antiferromagnet with open boundaries unpaired
spins should occur at both ends of the chain. Owing to them 
$\langle S_l^z\rangle$ decays exponentially as one moves towards the
center of the chain, if it is calculated for the lowest $S_T^z=1$ state,
which in the thermodynamic limit should become one of the components of the
fourfold degenerate ground state. This decay length was found to be
exactly the same as the correlation length in the bulk.

The analysis of critical half-odd-integer spin systems is more 
complicated. This is due to the existence of a marginally relevant
operator in the Hamiltonian of the spin-1/2 isotropic Heisenberg 
antiferromagnetic chain, which leads to important logarithmic corrections
to the power law decay of the spin-spin correlation function.
According to the renormalization group analysis based on effective
continuum models \cite{wzw,k9}, for large $l$ the correlation function 
behaves asymptotically as
\begin{equation}
  \langle S_i^z S_{i+l}^z \rangle  \sim a 
  \times \frac{(-1)^l}{l^\eta} \left[\log\left(\frac{l}{c}\right)\right]^\sigma
  -\frac{1}{4\pi^2l^2} \,,
\label{eq:logcor}
\end{equation}
where $a$ and $c$ are constants, $\eta = 1$ and $\sigma = 1/2$.
This form was confirmed numerically by Sandvik and Scalapino \cite{k11}
and recently by Hallberg {\em et al.} \cite{s12cikk} 
using a highly accurate DMRG calculation. 

If a local field is applied to a spin on the chain end, the loose edge spin
is oriented and a slowly decaying magnetization develops,
\begin{equation}   
   \langle S_l^z \rangle \propto \frac{h(-1)^l}{l^{x_m}}\/,
\label{eq:ordaff}
\end{equation}  
where $l$ measures the distance from the surface and $x_m$ is the bulk
scaling dimension of $S_l^z$. It follows from the scaling theory that
$x_m = \eta /2$. 

Recently Affleck \cite{affleck7} considered the critical exponent $\eta$ 
of the two-point correlation function and $x_m$ of the decay of the 
magnetization for the anisotropic spin-1/2 Heisenberg chain 
subjected to a transverse surface magnetic field. In the isotropic limit, his 
result for the spin-spin correlation function and the magnetization
confirms Eqs. (\ref{eq:logcor}) and (\ref{eq:ordaff}) with $\eta =1$
and $x_m = 1/2$.

In a finite chain, when the spins are oriented locally by an external field, 
this power-law decay will be modified depending on the boundary condition.
The assumption of conformal invariance \cite{cardy} allows us to predict
the profile of the magnetization and to extract the properties of the 
infinite system from those of finite samples. It has been shown by Burkhardt
and Xue\cite{burkhardt} that for Ising like models on a strip of width 
$L$, conformal invariance leads to the profile 
\begin{equation}
  \langle S_l^z \rangle \sim 
      \left[ \frac{L}{\pi} \sin \frac{\pi l}{L}\right]^{-x_m}
        F_{ab} \left( \cos \frac{\pi l}{ L}\right) \,,
\end{equation}
where the function $F_{ab}$ depends on the type of boundary applied
on the two sides $a$ and $b$.

In the case when the spins are fixed at $l=0$ and they are free at the 
other end, $l=L$, the decay of the magnetization is described by the 
simple form
\begin{equation}
  \langle S_l^z \rangle \sim 
      \left[ \frac{L}{\pi} \sin \frac{\pi l}{L}\right]^{-x_m}
      \left[ \cos \frac{\pi l}{2 L}\right]^{x_m^s} \,,
\label{eq:szl}
\end{equation}
where $x_m$ and $x_m^s$ are the bulk and surface scaling dimensions of 
the magnetization. Close to the free end the magnetization is 
proportional to
\begin{equation}
    \langle S_l^z \rangle \sim  (L-l)^{x_m^s-x_m} \,.
\label{eq:xs}
\end{equation}
In the Ising model, where $x_m^s=2x_m$, we get $(L - l)^{x_m}$. 

One of the aims of this paper is to check whether a similar
relationship holds for the Heisenberg model. We will see, that for 
half-odd-integer spin models with $S>1/2$, although they are critical, 
a new problem arises. As Ng \cite{ng} has shown, loose end spins show up 
not only in the valence-bond model of integer spins, where the end spin 
value is $S_{\rm end}= S/2$, but in half-odd-integer models as well with
$S_{\rm end}=(S-1/2)/2$. This was confirmed numerically by Qin {\em et al.} 
\cite{qin} using the DMRG method for spin chains with $S=1/2,1,3/2,2$ 
using open boundary condition. These end spins will lead to a 
completely different profile of the magnetization near the free boundary
for $S=1/2$ and $S > 1/2$.

Nevertheless conformal invariance can be used to check the scaling form 
of the spin-spin correlation function even for $S > 1/2$ half-odd-integer
spin models, as has been shown by Koma and Mizukoshi \cite{koma}. 
Their numerical results for $S=3/2$ are in agreement with the assumption 
that the correlations show asymptotically the same behavior as for $S=1/2$. 
This was later confirmed by Hallberg {\em et al.} \cite{s32cikk} using 
very extensive DMRG calculations, showing that Eq. (\ref{eq:logcor}) is 
valid for the $S=3/2$ chain, too, with $\eta=1$ and $\sigma=1/2$. 

\section{Formulation of the problem}

In the present paper we will study the response of a 1-D spin-S Heisenberg 
model to a small magnetic field $h$ applied locally. The Hamiltonian of the 
system is written in the form
\begin{equation}
  {\cal H}=\sum_i J\vec S_i \vec S_{i+1} + {\cal H}_{\rm local} \,,
\label{eq:ham}
\end{equation}
where the first term is the usual first neighbor Heisenberg model
with antiferromagnetic interaction and the second term denotes the coupling 
of the field to a spin. In what follows we will take $J=1$ and measure the 
field relative to this coupling.

As mentioned before, effective edge spins may appear in the vicinity of 
the boundaries even without external field. In order to separate the 
magnetization induced by the field and the effect of the edge spins, the 
magnetic field should be applied far from the ends. In our calculations
we will apply it on the central site of the chain. 

In a finite field the ground state does not necessarily belong to the 
singlet, $S_T=0$, spin sector and the spin quantum number
of the ground state may change as a function of the chain length. This 
may lead to complications in the finite-size scaling procedure. 
These difficulties can be eliminated by applying the magnetic field on
two neighboring sites with opposite strength, in which case the ground
state remains always in the singlet sector. We have therefore used
the following expression for the local Zeeman term: 
\begin{equation}
  {\cal H_{\rm local}}=h S^z_{N/2}-hS^z_{N/2+1}\/.
\end{equation}

Since we are interested in the response to weak fields, we tried to remain
in the regime where the local magnetization is proportional to
the applied field. It turned out that 
for integer spins, linearity holds at least up to 
$h<0.25$.  
For half-odd-integer spins
this regime was found to be much smaller, thus $h<0.1$ fields 
were used for the $S=1/2, 3/2$ chains. 

Our calculations were performed using the infinite-lattice 
method of DMRG. First a chain with $N-2$ sites is built up without field.
In the case of open boundary condition we have used both the 
spin-reversal and the left-right reflection symmetries in order to reduce 
the size of the Hilbert-space of the superblock configuration. The field
was introduced in the last step of the infinite-lattice method on the
two sites between the left and right blocks. The number of block states 
kept in the calculation varied between 100 and 200. The truncation error 
was better than $10^{-7}-10^{-8}$. As an indication of the absolute error 
of our calculation, we point out that in zero external field in the 
$S_T^z=0$ spin sector the largest value of the site magnetization was 
$10^{-5}$.

Since boundary effects are expected to influence the finite-size 
calculations we have considered both open and periodic boundary conditions.
In some cases real $S=1/2$ spins were attached to the end of the $S=1$
and $S=3/2$ chains, further reducing the contribution of the free end spins.

\section{Numerical results}

In this section we present the results of our numerical calculations.

\subsection{The S=1 case}

Chains up to $N=120$ sites were considered. This allows us to 
separate the boundary effects and the local magnetization induced by
the field. Fig.\ \ref{fig:s1obc} shows the absolute value of 
the site magnetization calculated with open boundary condition in the 
ground state which is a total spin singlet ($S_T=0$) state, for two values 
of the field applied on the central sites. The striking result of this 
calculation is the appearance of the end spins even in the $S_T^z=0$ sector
for arbitrarily small finite field. It is also clearly seen in the figure
that in the region around the perturbed sites, where the magnetization decays, 
the amplitude is proportional to the strength of the local field, while 
close to the free boundary the magnetization is independent of the field. 
As a comparison Fig.\ \ref{fig:s1obc} also shows the site magnetization 
calculated for the lowest lying state in the $S_T^z=1$ spin sector in the 
absence of field. 

Since the system is always gapped, an exponential decay is obtained in all 
cases, both near the edge and around the perturbed site. The decay length 
was determined, following White and Huse \cite{s1cikk} from the quantity
\begin{equation}
    \xi_i\equiv -1/\ln [-C(i)/C(i-1)] \,,
\label{eq:zetai}
\end{equation}
where $C(i)=\langle S_i^z \rangle$, since it gives a clear indication of 
the region where the asymptotic behavior sets in. From the semi logarithmic 
plot used in the figure it is clear that near the edge the slopes of the 
curves obtained for different field values are the same. When $M=120$
states are kept in the DMRG we have found $\xi_{\infty}\simeq 5.85(5)$, 
reasonably close to the known best result for the decay length. 
 
The analysis can be done in the same way for the decay length around the 
central perturbed site using
\begin{equation}
    \xi_l\equiv -1/\ln [-C(l)/C(l-1)] \,,
\label{eq:zetal}
\end{equation}
where $l$ is measured from the perturbed site, from the middle of the chain. 
The same value is obtained as above. These short decay lengths ensure that 
the effects of the edge spins and of the locally applied field could 
clearly be separated.

To further clarify the origin of the nonvanishing magnetization at the 
boundary, real $S=1/2$ spins were attached to the ends of the chain with 
adjustable $J_1=J_{N-1}\equiv J_{\rm end}$ couplings. Fig. \ref{fig:s1b12} 
shows the site dependence of the magnetization on a semi logarithmic plot 
calculated for $J_{\rm end}=1.5$. It is compared to the situation when no 
extra spins are attached. i.e., $J_{\rm end}=0$.
 
From these results it is found that the effect of the edge spin can be 
eliminated by the extra spin, while the magnetization is not modified 
far from the edge. The fit of the curves obtained with $M=180$ states kept 
leads to $\xi_{\infty}=5.95(3)$, a result closer to the expected value, than
that obtained for $M=120$. The accuracy can be improved by fitting the 
quantities $\xi_{\infty}(M)$ calculated for various values of $M$ as a 
function of $1/M$. In the $M\to\infty$ limit the value 
$\xi_{\infty}\simeq 6.05(5)$ is found. 

To complete the analysis, let us consider the effect of the field on the
spin-spin correlation function. In the absence of magnetic field the 
two-point correlation function is usually calculated in the DMRG method 
by taking the two sites symmetrically with respect to the center of the 
chain. On the other hand, in the presence of the magnetic field acting on 
the site $i=N/2$, the correlation function $\langle S_i^z S_{i+l}^z\rangle$ 
was calculated for $i=N/2$. Employing again Eq.\ (\ref{eq:zetal}) but with 
$C(l)=\langle S_i^z S_{i+l}^z \rangle$, and fitting the correlation 
functions obtained for various values of $M$ and at different field values 
in the $0<h<0.25$ interval, the value $\xi_{\infty}=5.92(5)$ is obtained 
for the correlation length. 

Within the numerical accuracy the numbers obtained for the decay length,
agree with those obtained by White and Huse \cite{s1cikk}. We can 
therefore conclude that the decay length and the correlation length of the 
two-point correlation functions are identical and field independent. For 
weak fields, where linear response is valid, there is only one length 
scale in the system determined by the gap.

In order to have a further check, similar calculations were done also with 
periodic boundary condition. Our results for $M=20-180$ block states are 
plotted in Fig.\ \ref{fig:s1pbc}. For small $M$ values the magnetization 
curves are not symmetric and a minimum is observed in the curves. This is 
in fact due to the truncation procedure in the DMRG wave function. As the 
number of the block states is increased the minimum disappears and the 
decay of the magnetization becomes more symmetric. Carrying out again 
the $\xi (M)$ extrapolation as a function of $1/M$, the value of the 
correlation length scales to $\xi_{\infty}=6.0(1)$ in the $M\to\infty$ limit. 

It is worth mentioning that the asymmetry mentioned above is related to a 
similar asymmetry in the local energy $\langle \vec S_i \vec S_{i+1}\rangle$ 
inside the blocks used in the DMRG method, if a relatively small number of 
block states is kept. This asymmetry provides therefore a better 
information about the real error than the truncation error \cite{white} 
defined in the standard way.

\subsection{The S=1/2 case}

The analysis of the critical half-odd-integer spin system is more complicated
due to the logarithmic corrections. Our calculation performed on chains with 
some $100$ sites in the interval $0<h<0.1$, where the magnetization is a 
linear function of the field. 

First we show our results for the decay of the magnetization. The absolute 
value of the site magnetization calculated in the $S_T^z=0$ spin sector 
with $M=180$ block states and open boundary condition for $h=0.02$ and
$h=0.05$ are shown in Fig.\ \ref{fig:s12obc}. Both curves were obtained
for the case when the field was applied on two central sites in the 
configuration $h_{N/2}=-h_{N/2+1}$. In order to avoid complications coming
from the $1/l^2$ term of Eq.\ (\ref{eq:logcor}), only the results on even 
sites are shown.

Using a fit to the form in Eq.\ (\ref{eq:ordaff}) by measuring the 
distance $l$ from the center of the chain we have found $A\simeq 0.37(1)$ 
and $x_m =0.49(3)$. Since in this case the magnetization shows a 
downward curvature near the free end, we assumed there the form
(\ref{eq:xs}). We have got a good fit with $x_m^s - x_m = 0.49(2)$. 
Basically the same results were obtained when a single site was
perturbed and also for other, larger values of the field.
Thus we find that like in the Ising model, the relationship
$x_m^s=2 x_m$ holds for the Heisenberg model.
 
The effect of the field on the spin-spin correlation function has also been
investigated. Following the proposal of Ref.\ [10] we have 
considered the average value 
\begin{equation}
   \overline{\omega} = \frac{1}{4}[ \omega(l-1,N)+ 2 \omega (l,N) + 
     \omega(l+1,N)]
\label{eq:omega}
\end{equation}
to remove the $l^{-2}$ correction of Eq. (\ref{eq:logcor}), where 
$\omega(l,N)=(-1)^l\langle S_{i+l}^z S_i^z\rangle$. Our data obtained 
for a chain with $N=100$ sites with $M=180$ block states for zero and 
finite ($h=0.25$) field are plotted in Fig.\ \ref{fig:s12cor}. Eq. 
(\ref{eq:logcor}) gives a good fit both without and with field, confirming 
that apart from logarithmic corrections the leading term in the 
correlation function goes as $1/l$ even in the non-linear regime. This 
result also provides us with a test of the accuracy of our result when 
compared to those in Ref. [10].

As a further check we have considered the same problem with periodic
boundary condition. The deviation of the local energy from the value obtained
at the $i=N/2$ site was of the order of $10^{-2}$ in the interior of the
block from which the system is built up in DMRG, implying that for better 
accuracy more block states should be used. In spite of this error we have 
found $x_m=0.48(5)$ for the decay of the magnetization and $\eta \simeq 1$ 
in the spin-spin correlation function in agreement with the result 
obtained recently by Affleck \cite{affleck7}.

\subsection{The S=3/2 case}

It follows from our previous discussion that open boundary condition
usually gives better results, in this case, however, we have to face again 
the problem of edge spins.

The result of the DMRG calculation performed with open boundary condition 
and $h=0.05$ for various values of the block states is plotted in 
Fig.\ \ref{fig:s32obc}. Two things are apparent: the 
contribution of the end spins causes a nonvanishing value of the site 
magnetization close to the chain ends and an even-odd oscillation is
present even if the absolute value of the site magnetization is shown.
Since the system is critical, the magnetization falls off algebraically.
Therefore the effect of the field and 
of the end spins cannot be separated for chains for which the calculation
could be done. In order to remove the contribution of the end spins we have 
attached real $S=1/2$ spins to the chain ends. The site magnetization
is shown in Fig.\ \ref{fig:s32b12}. The fit to these data gives
for the decay exponent $x_m=0.50(2)$, the same value as obtained
for the $S=1/2$ case. Since according to the result of Ref.\ [16] 
the exponent of the decay of the spin-spin correlation function is $\eta=1$, 
the relation $x_m = \eta /2$ holds for $S=3/2$, too.

The calculation of the profile for free end spins would need at least a few 
thousand block states as in Ref.\ [16].

\section{Conclusion}

In the present paper we have considered spin-S antiferromagnetic
Heisenberg spin chains in a locally applied oppositely oriented magnetic 
field on two sites in the center of the chain, leaving the boundary 
spins free.  

For the spin-1/2 Heisenberg chain it has been found that the linear response 
of the system holds only for extremely small perturbations. The exponent 
describing the decay of the site magnetization was found to be $x_m=1/2$ 
around the locally applied field, while $\eta =1$ is found for the 
spin-spin correlation function independently of the field thus confirming 
numerically, at least in the isotropic point, the analytical result obtained 
recently by Affleck\cite{affleck7}. Near the boundary the decay of the site 
magnetization changes character and disappears at the edge. Assuming the 
profile predicted by conformal invariance, $\langle S_l^z \rangle \propto 
(L-l)^{x_m^s -x_m}$ with $x_m^s = 2 x_m$, very much like in the Ising model.

For the integer spin case it was shown that linear response theory is
satisfied for stronger perturbations, too. The field applied in the center 
of the chain, however weak it is, will orient the end spins and the site 
magnetization calculated in the $S_T^z=0$ ground state shows a dramatic 
increase near the ends. We have confirmed that there exists a single 
length scale in the system, related to the finite gap. The various bulk 
and surface quantities fall off exponentially  with the same correlation 
length. In particular, we have shown that the correlation length in the 
two-point correlation function is identical to the decay length of the
magnetization and that they are field independent. 

Using periodic boundary condition we have pointed out that a better measure
of the accuracy of the DMRG procedure than the truncation error could be
obtained from looking at the deviation of the local energy from its mean value.

Finally the relationship $x_m = \eta/2$ has been shown to hold
for the spin-3/2 Heisenberg chain, as well. Although this model has the 
same bulk critical exponents as the $S=1/2$ Heisenberg model, the
magnetization profile is very different, resembling much more that of the
$S=1$ model, indicating that indeed effective $S=1/2$ spins appear
at the edges. Unfortunately, our limited computational resources did 
not allow us to determine the precise form of the profile.   

\section{Acknowledgments}

This research was partially supported by the Hungarian Research Fund
(OTKA) Grant No.\ 15870. We acknowledge I. Peschel, D. J. Scalapino and 
T. Ziman for useful discussions, the U.S.-Hungarian Joint Fund and the 
French Ministry of Research and Technology for supporting our stays in 
Santa Barbara and at the Institut Laue Langevin (Grenoble), respectively.

\newpage

\begin{figure}
\caption{The absolute value of the site magnetization calculated in the 
$S_T^z=0$ spin sector of the ground state for an $S=1$ spin chain with 
$N=120$ sites for two values of the locally applied field. For comparison 
the magnetization obtained in the lowest lying state of the $S_T^z=1$ spin 
sector without field is also shown.}
\label{fig:s1obc}
\end{figure}

\begin{figure}
\caption{The same as Fig. 1. for one value of the field for the case when 
an extra $S=1/2$ spin is coupled to the chain ends with the exchange 
coupling $J_{\rm end}$. Only the values for the left half of the chain 
are plotted.} 
\label{fig:s1b12}
\end{figure}

\begin{figure}
\caption{The same as Fig. 1. for one value of the field but for periodic 
boundary condition, for various values of the number of states kept. The 
straight lines are fits with exponential decay.}
\label{fig:s1pbc}
\end{figure}

\begin{figure}
\caption{The site magnetization of a $S=1/2$ spin chain obtained in the 
$S_T^z=0$ spin sector with open boundary condition. $N=120$, M=$180$ 
$h=0.02$ and $h=0.05$. The solid lines correspond to our fits explained 
in the text.}   
\label{fig:s12obc}
\end{figure}

\begin{figure}
\caption{The averaged spin-spin correlation function of the $S=1/2$ spin chain 
obtained in the $S_T^z=0$ spin sector with open boundary condition in the 
absence of the field and for $h=0.25$. The solid lines correspond to our 
fits explained in the text.}   
\label{fig:s12cor}
\end{figure}

\begin{figure}
\caption{The site magnetization of $S=3/2$ magnetic chain with $N=60$ sites
calculated for various values of $M$ with open boundary condition.} 
\label{fig:s32obc}
\end{figure}

\begin{figure}
\caption{Same as Fig. 6. but with real $S=1/2$ spins attached to the 
chain ends.}
\label{fig:s32b12}
\end{figure}

\end{document}